\documentclass[12pt,twoside]{article} 

\usepackage[mathscr]{euscript}

\date{14-3-2013} 

\begin{document}

\centerline{} 

\centerline{} 

\centerline {\Large{\bf CHSH is not supported with supermartingale statistics.}} 

\centerline{} 

%\centerline{\Large{\bf Title Second Line}} 

\centerline{} 

\centerline{\bf {J.F. Geurdes}} 

\centerline{} 

\centerline{C. vd Lijnstraat 164} 

\centerline{2593 NN Den Haag} 

\centerline{Netherlands} 

%\centerline{Address of Author1 forth line} 

\newtheorem{Theorem}{\quad Theorem}[section] 

\newtheorem{Definition}[Theorem]{\quad Definition} 

\newtheorem{Corollary}[Theorem]{\quad Corollary} 

\newtheorem{Lemma}[Theorem]{\quad Lemma} 

\newtheorem{Example}[Theorem]{\quad Example} 

\centerline{}

%{\footnotesize Copyright $\copyright$ 20xx Author 1 and Author 2. This is an open access article distributed under the Creative Commons Attribution %License, which permits unrestricted use, distribution, and reproduction in any medium, provided the original work is properly cited.}

\begin{abstract}It is demonstrated that the supermartingale statistics approach of Gill to the CHSH contrast contains a physical (and statistical) unrealistic assumption.
\end{abstract} 



\section{Introduction}The CHSH statistic arises in discussions of locality and causality with additional parameters in quantum mechanics. The argument starts with Bell's correlation expression \cite{1}
\begin{equation}\label{Cor1}
E(a,b)=\int_{\lambda \in\Lambda} \rho_{\lambda}A_{\lambda}(a)B_{\lambda}(b)d\lambda
\end{equation}
Here $\rho_{\lambda}$ is the probability density function of the hidden variables $\lambda \in \Lambda$. The $A_{\lambda}(a)$ and $B_{\lambda}(b)$  are measurement functions that project in $\{-1,1\}$.The $a$ and $b$ refer to the setting parameters. In CHSH studies $a$ and $b$ refer to labels in $\{1,2\}$ that, in turn, refer to unit parameter vectors. From (\ref{Cor1}) it can be derived that a contrast $S=E(1,1)-E(1,2)-E(2,1)-E(2,2)$ exist such that $|S| \leq 2$. This is called the CHSH contrast.   
\\\\
In \cite{2} and \cite{3} it is argued that violation with CHSH with local hidden variables is impossible because the CHSH statistics follows a supermartingale structure. This comprises Gill's formulation of Bell's fifth position: "It will be impossible to simulate a CHSH violation using strict Einstein locality with a computer model". Local hidden variables are additional parameters that in a local manner explain the correlation between distant particles. \\\\ In the present paper the supermartingale argument against local hidden causality will be found to be in conflict with observation. 

\section{Preliminary definitions counting and response probability} In the first place we follow Gill in his definition of a finite elementary statistical counting measure.
\subsection{Counting measure}
\begin{equation}\label{CountM1}
\Delta_n(a,b)=1\{(X_n,Y_n)=(x_n,y_n)\, :\, x_n=y_n\}1\{(A_n,B_n)=(\mathbf{a},\mathbf{b})\}
\end{equation}
The $(\mathbf{a},\mathbf{b})$ refer to the physical unit parameter vectors $|\mathbf{a}|=1$ and $|\mathbf{b}|=1$. Later the notation will be replaced with labels $(a,b)$. We see that $\Delta_n(a,b) \in \{0,1\}$. The pair $(x_n,y_n) \in \{-1,1\}^2$ refer to the outcome of the measurements. $X_n$ at Alice's measuring instrument and $Y_n$ at Bob's. $1\{(X_n,Y_n)=(x_n,y_n)\, :\, x_n=y_n\}$ is unity when $ x_n=y_n$ and zero otherwise. The uppercase letters indicate random variables. The lowercase letters refer to the value of random variables. The index n is the trial number. If there are e.g. $N$ trials then e.g. the number of equal measurements at setting pair $(\mathbf{a},\mathbf{b})$ is $N^{=} (\mathbf{a},\mathbf{b})$. This is
\begin{equation}\label{Neq}
N^{=}(\mathbf{a},\mathbf{b})=\sum_{n=1}^{N} \Delta_n(\mathbf{a},\mathbf{b})
\end{equation}
From the definition of the $1\{(A_n,B_n)=(\mathbf{a},\mathbf{b})\}$ we can then derive that $N(\mathbf{a},\mathbf{b})=\sum_{n=1}^{N}1\{(A_n,B_n)=(\mathbf{a},\mathbf{b})\}$. Hence, there are $N^{\neq}(\mathbf{a},\mathbf{b})=N(\mathbf{a},\mathbf{b})-N^{=}(\mathbf{a},\mathbf{b})$ unequal, i.e. $x_n \neq y_n$ measurements when 'no measurements are lost'. The product moment $\hat{\rho}(\mathbf{a},\mathbf{b})=\hat{\rho}(a,b)$ is defined by
\begin{equation}\label{ProdMom}
\hat{\rho}(a,b )=\frac{ N^{=}(a,b) - N^{\neq}(a,b) }{ N(a,b) }=\frac{2N^{=}(a,b)}{N(a,b)}-1
\end{equation}
Here we will use 'labels' that refer to parameter vectors.  
\subsection{CHSH transform}
Let us defined a CHSH weight based on (\ref{CountM1}) like $\mathcal{C}_{(1,1)}(\hat{\rho}(a,b))$ such that
\begin{equation}\label{Transf}
\mathcal{C}_{(1,1)}(\hat{\rho}(a,b))=\hat{\rho}(1,1)-\hat{\rho}(1,2)-\hat{\rho}(2,1)-\hat{\rho}(2,2)
\end{equation}
 With this notation let us subsequently introduce $\Delta_n=\mathcal{C}_{(1,1)}(\Delta_n(a,b))$. We have $\Delta_n \in \{-1,0,1\}$. For completeness, the outcome zero occurs when $(a,b)$ has been selected at the $n$-th trial but $x_n\neq y_n$. In a proper experiment the probability of this event will be low. Let us subsequently define the random variable $Z_N$ as a sum of the $\Delta_n$ for $n=1,2,...  $.
\begin{equation}\label{Z1}
Z_N = \sum_{n=1}^{N} \Delta_n
\end{equation}
When $N$ is the total number of trials and is large enough then we can safely assume that $N(a,b)\approx N/4$ then the following relation between $Z_n$ and $S$ can be obtained: $Z_N \approx (S-2)N/8$. The aim of Gill's argument is to show that $Z_n$ finally decreases in $n=1,2,....N$ and wil likelyl be $\le 0$ in the end. 

\subsection{Expectation value \& probability}
Looking at the expected behavior of $Z_n \sim \,\,<0$ for increasing $n$, Gill introduces the supermartingale assumption.
\begin{equation}\label{Super}
E(Z_{n+1}\,|\,Z_1=z_1,Z_2=z_2,....,Z_n=z_n)\leq z_n
\end{equation}  
and tries to show with a Markov equation that the probability of $\max(Z_n) >0$ decreases with increasing $n$.
\\\\
In the present argument of our paper we will be in need of conditional probabilities in a discrete probability space. Let us concentrate on the random variables $\Delta_n$. A conditional density, equal to conditional probability in the discrete case, for the $\Delta_n$ is
\begin{equation}\label{CndDens}
f_{\Delta_{n+1}\,|\,\Delta_1,....\Delta_n}(\delta_{n+1}|\delta_1,...,\delta_n)=\frac{f_{\Delta_1,....\Delta_{n+1}}(\delta_1,...,\delta_n,\delta_{n+1})}{\sum_{\delta_{n+1}\in\{-1,0,1\}}f_{\Delta_1,....\Delta_{n+1}}(\delta_1,...,\delta_n,\delta_{n+1})}
\end{equation} 
In addition, we also will be in need to derive a density for $\{\delta_1,...,\delta_n\}$ from a density of $\{\delta_1,...,\delta_n,\delta_{n+1}\}$ with \cite{4}
\begin{equation}\label{ReducedProb}
f_{\Delta_1,....,\Delta_n}(\delta_1,...,\delta_n)=\sum_{\delta_{n+1}\in\{-1,0,1\}}f_{\Delta_1,....\Delta_{n+1}}(\delta_1,...,\delta_n,\delta_{n+1})
\end{equation}
We assume that the probabilities can be small but in all cases non-zero. This appears a reasonable assumption for this type of research.

\section{Results and Discussion} These are the main results of the paper.
From equation (\ref{Z1}) the supermartingale condition in (\ref{Super}) can be transformed in terms of $\Delta_n$, with $n=1,2,...N$. In terms of $\Delta$ the expectation can be rewritten like
\begin{eqnarray}\label{SuperEx}
\begin{array}{cc}
E(\Delta_{n+1}\,|\,\Delta_1=\delta_1,\Delta_2=\delta_2,....,\Delta_n=\delta_n)=\\\\
~~~~~~~~~~~~~~~~~~\sum_{\delta_{n+1}\in\{-1,0,1\}}\delta_{n+1}f_{\Delta_{n+1}\,|\,\Delta_1,....\Delta_n}(\delta_{n+1}|\delta_1,...,\delta_n)
\end{array}
\end{eqnarray}
The following theorem can be easily proved. It shows the transformation of a $Z$ supermartingale into a $\Delta$ supermartingale.

\begin{Theorem} The $Z$ form of the CHSH supermartingale can be transformed into a $\Delta$ form as. 
\begin{equation}
E(\Delta_{n+1}\,|\,\Delta_1=\delta_1,\Delta_2=\delta_2,....,\Delta_n=\delta_n)\leq 0
\end{equation} 
\end{Theorem} 
This follows quite easily from equation (\ref{Z1}) and equation (\ref{Super}) because the expectation is a linear function of a sum of random variables. Note: $z_n=\sum_{m=1}^{n}\delta_m$ and the value of $\Delta_m$ in a conditioned expectation where $\Delta_m=\delta_m$ is of course $\delta_m$. Hence $z_n$ on the right hand and the left hand side drop off when $Z_{n+1}=z_n +\Delta_{n+1}$. A consequence of this theorem is that 
\begin{equation}\label{Conseq1}
f_{\Delta_{n+1}\,|\,\Delta_1,....\Delta_n}(-1|\delta_1,...,\delta_n)\geq f_{\Delta_{n+1}\,|\,\Delta_1,....\Delta_n}(\delta_{n+1}|\delta_1,...,\delta_n)
\end{equation}
for $\delta_{n+1} \in \{-1,0,1\}$. From equation (\ref{CndDens}) it follows that with the unconditioned density we must have
\begin{equation}\label{Conseq2}
f_{\Delta_1,..\Delta_{n+1}}(\delta_1,..\delta_{n},-1)\geq f_{\Delta_1,..\Delta_{n+1}}(\delta_1,..\delta_n,\delta_{n+1})
\end{equation}
In order to see a steady decline in $\{Z_n\}_{n=1}^N$ i.e. in $\{\Delta_N\}_{n=1}^N$, the need is there for a strict supermartingale. Moreover, because in proper experimentation the $\delta_n=0$ has a low probability we can derive,from (\ref{Conseq2}), for a strict supermartingale that 
\begin{equation}\label{Conseq3}
f_{\Delta_1,..\Delta_{n+1}}(\delta_1,..\delta_n,-1)>f_{\Delta_1,..\Delta_{n+1}}(\delta_1,..\delta_n,1)>f_{\Delta_1,..\Delta_{n+1}}(\delta_1,..\delta_n,0)
\end{equation}
From (\ref{ReducedProb}) we may also derive that 
\begin{equation}\label{sumProb}
f_{\Delta_1,....\Delta_{n+1}}(\delta_1,...,\delta_n,-1)\geq \sum_{\delta_{n+2}\in\{-1,0,1\}}f_{\Delta_1,....\Delta_{n+2}}(\delta_1,...,\delta_n,\delta_{n+1},\delta_{n+2})
\end{equation}
Hence, $f_{\Delta_1,....\Delta_{n+1}}(\delta_1,...,\delta_n,-1)>f_{\Delta_1,....\Delta_{n+2}}(\delta_1,...,\delta_n,\delta_{n+1},-1)$ whenall  probabilities are assumed unequal to zero..

\section{Conclusion} 
The conclusion from the previous analysis is that in order to have a a (strict) supermartingale the probability of the $\delta_i =-1$ response declines with the increase of the trial number. This is so because: $f_{\Delta_1,....\Delta_{n+1}}(\delta_1,...,\delta_n,-1) > f_{\Delta_1,....\Delta_{n+2}}(\delta_1,...,\delta_n,\delta_{n+1},-1)$. The decline of probability to see a $\delta_i=-1$ is physically unrealistic.
\\\\
The strict matringale is (every now and then) in the series of trials $n=1,2,...$ necessary in order to warrant a steady decline of $\Delta_n$ and hence of $Z_n$. Hence, Gill's supermartingale argument and the subsequent application of Hoefding's inequality is based upon an unrealistic assumption. The assumption is unrealistic in a physical and probabilistical sense. Hence, there is no plausibile case for a fifth position in the dispute on locality and causality in quantum mechanics. This finding supports previous results of the author \cite{5}, \cite{6} and \cite{7}

{\bf Received: Month 03, 2013}

\end{document}